\documentclass[10pt,pra,twocolumn,floatfix]{revtex4}
\usepackage{amsfonts}
\usepackage{amssymb}
\usepackage{amsmath}
\usepackage[dvips]{graphicx}

\begin{document}

\title{Compact sub-kilohertz low-frequency quantum light source based on four-wave mixing in cesium vapor}

\author{Rong Ma$^{1}$}
\author{Wei Liu$^{1}$}
\author{Zhongzhong Qin$^{1,2}$}
\email{zzqin@sxu.edu.cn}
\author{Xiaolong Su$^{1,2}$}
\author{Xiaojun Jia$^{1,2}$}
\author{Junxiang Zhang$^{1,2,3}$}
\author{Jiangrui Gao$^{1,2}$}
\affiliation{$^1$State Key Laboratory of Quantum Optics and Quantum Optics Devices,
Institute of Opto-Electronics, Shanxi University, Taiyuan 030006, People's
Republic of China\\
$^2$Collaborative Innovation Center of Extreme Optics, Shanxi University,
Taiyuan, Shanxi 030006, People's Republic of China\\
$^3$Department of Physics, Zhejiang University, Hangzhou, 310027, People's Republic of China\\
}

\begin{abstract}
Using a nondegenerate four-wave mixing (FWM) process based on a double-$\Lambda$ scheme in hot cesium vapor, we demonstrate a compact diode-laser-pumped quantum light source for the generation of quantum correlated twin beams with a maximum squeezing of 6.5 dB. The squeezing is observed at a Fourier frequency in the audio band down to 0.7 kHz, which is the first observation of sub-kHz intensity-difference squeezing in an atomic system so far. Phase-matching condition is also investigated in our system, which confirms the spatial-multi-mode characteristics of the FWM process. Our compact low-frequency squeezed light source may find applications in quantum imaging, quantum metrology, and the transfer of optical squeezing onto matter wave.
\end{abstract}

\maketitle

Quantum light sources, such as squeezed state and continuous variable (CV) Einstein-Podolsky-Rosen entangled state, are fundamental ingredients for applications in future quantum technologies, such as quantum metrology \cite{QuMetrology}, quantum computation \cite{QuComputation}, and quantum information processing \cite{BraunsteinRMP}. These applications introduce specific requirements on the quantum light sources. For instance, squeezed-light enhancement of Laser Interferometer Gravitational-Wave Observatory (LIGO)  has been experimentally demonstrated \cite{GravitationalWave1,GravitationalWave2,GravitationalWave6}, which is realized by injecting squeezed light into the interferometer's normally unused port to replace relevant vacuum fluctuations following the proposal by Caves \cite{CavesPRD}. Recently, a new scheme is proposed to improve the sensitivity of LIGO by making use of CV entangled states \cite{GravitationalWave3}. In recent years, attention has also been brought to the problem of manipulating cold atomic ensembles with quantum states in order to produce nonclassical matter waves \cite{MatterWave1,MatterWave2,MatterWave3}. For all of the above applications, squeezing at a Fourier frequency in the audio band (to frequencies above $\approx$10 Hz, up to about 10 kHz) is required. On the other hand, for quantum states to be used as quantum information carriers interacting with other systems, such as atoms and solid-state systems, the central frequency and linewidth of the quantum states must match the transitions of matter to ensure an efficient coupling between light and matter \cite{LukinRMP}.

The standard technique to generate squeezed state and CV entangled states is by parametric down-conversion in a nonlinear crystal, with an optical parametric oscillator or optical parametric amplifier \cite{LauratOL,JiaOE,VillarPRL,SuOL,JingPRA}. Squeezed vacuum state with a maximum 15 dB noise reduction has been achieved in this way \cite{SchnabelPRL}, and squeezing with a Fourier frequency in audio band down to 1 Hz have also been reported by several groups \cite{SchnabelNJP,PKLamLowFreq}. However, the central frequency and linewidth of the generated squeezed states usually do not naturally match the transitions of matter, such as atoms and solid-state systems, which limits their applications in light-matter interactions.

A promising alternative method is to produce quantum states with an atomic system directly. Actually, the first experimental demonstration of a squeezed state of light was realized using a four-wave mixing (FWM) process in sodium vapor \cite{SlusherPRL}. In the low-frequency region, a squeezed vacuum state with a Fourier frequency down to 100 Hz is generated based on the polarization self-rotation effect in atomic vapor \cite{Magnetometer1}. Recently, it was shown that quantum correlated twin beams can be generated based on the FWM process in a double-$\Lambda$ scheme in hot rubidium vapor \cite{LettPRA,LiuOL,QinOL,GlorieuxArxiv,JasperseOE} and cesium vapor \cite{MaPRA}. This system has proven to be very successful for a variety of applications such as the generation of multiple quantum correlated beams \cite{QinPRL,QinAPL}, entangled images \cite{EntangledImages}, high purity single photons \cite{QinLight}, as well as optical qubits \cite{TravisOL}, the tunable delay of EPR entangled states \cite{TunableDelay}, the realization of a SU(1,1) nonlinear interferometer \cite{SU11NC}, and the ultrasensitive measurement of microcantilever displacement below the shot-noise limit \cite{PooserDisplacement}. The lowest Fourier frequency where intensity-difference squeezing is observed in this scheme is 1.5 kHz and 8.0 kHz for a Ti:sapphire-laser-pumped system \cite{LiuOL} and a diode-laser-pumped system \cite{QinOL}, respectively. 

In this Letter, we demonstrate a compact diode-laser-pumped system for the generation of intensity-difference squeezing down to 0.7 kHz with a maximum squeezing of 6.5 dB based on the FWM process in hot cesium vapor. The achievement of sub-kHz intensity-difference squeezing can be mainly attributed to two factors. On one hand, the one-photon gain resonance in cesium vapor is much broader than that in rubidium vapor. On the other hand, the mechnical stability of the compact quantum light soure and the electronic noise of the detection system are optimized. Especially, the diode laser is placed on an independent, insulated and actively damped optical table to reduce the coupling with the environment. The phase matching condition of the FWM process in our system is also investigated. Compared with bulky and expensive Ti:sapphire lasers used in most of the above experiments \cite{EntangledImages,QinLight,TravisOL,TunableDelay,SU11NC}, diode lasers are much cheaper, more compact, and easier to operate. The presented results provide a useful compact quantum light source for applications in quantum imaging and quantum metrology.

\begin{figure}[t]
\includegraphics[width=9cm]{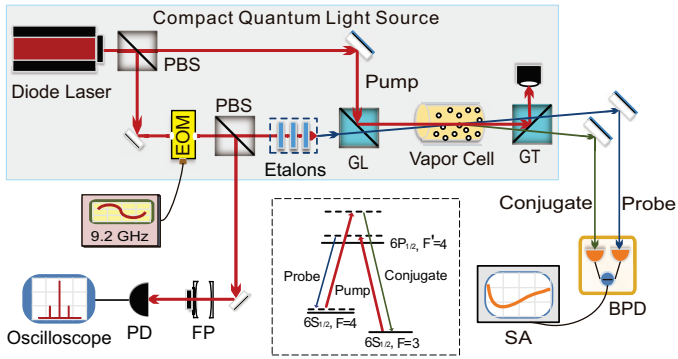}
\caption{(Color online) Experimental setup for generating and detecting quantum correlated twin beams. PBS: polarization beam splitter; EOM: electro-optic modulator; FP: Fabry-Perot interferometer; PD: photodetector; GL: Glan-laser polarizer; GT: Glan-Thompson polarizer; BPD: balanced photodetector; SA: spectrum analyzer. Inset: Double-$\Lambda$ scheme in the $D_{1}$ line of $^{133}$Cs.}
\end{figure}

As shown in Fig. 1, the FWM process relies on a double-$\Lambda$ scheme in which two pump photons are simultaneously converted to one probe photon and one conjugate photon. The process is coherent, and the intensities of the twin beams, i.e., the probe and conjugate beams are highly correlated with an intensity-difference noise below the corresponding shot noise limit (SNL). Our experiment begins with a diode laser which is tuned about 1.6 GHz to the blue of the $^{133}$Cs (895 nm, 6S$_{1/2}$, F=3$\rightarrow$6P$_{1/2}, F'=4$) with a total power of about 900 mW. The diode laser is placed on an isolated optical table to minimize the coupling to the environment, and the output of the laser is coupled to another optical table for experiment through a single-mode fiber (not shown in Fig. 1), which also serves as a spatial filter in the meanwhile. The laser beam is split into two beams by a polarization beam splitter (PBS). One of the beams serves as the pump beam, and the other beam passes through an electro-optic modulator (EOM) to produce optical sidebands at $\pm$9.2 GHz from the carrier frequency. The modulation efficiency of the EOM is monitored by a scanning Fabry-Perot (FP) interferometer. Three successive temperature-stabilized etalons (with a thickness of 7 mm, 7 mm, and 3 mm, respectively)  are used to select the probe frequency component ($-1^{st}$ order sideband) from the carrier and two sidebands. It has been shown that the frequency and phase-difference stability between the pump and probe beams is very good with this method\cite{MaPRA}.

\begin{figure}[t]
\includegraphics[width=8.5cm]{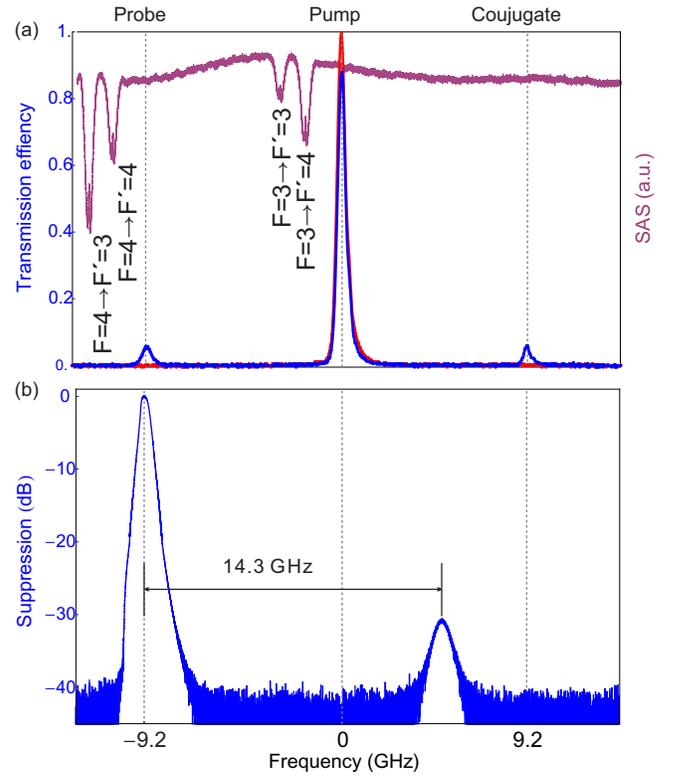}
\caption{(Color online) (a) Transmission spectra of the scanning FP interferometer when the EOM modulation is off (red curve) and on (blue curve), respectively. The purple curve shows the saturated absorption spectrum (SAS) of the $^{133}$Cs $D_{1}$ line as the frequency reference. (b) Background suppression compared to signal of three successive temperature-stabilized etalons. 14.3 GHz corresponds to the free spectrum range of the 7 mm etalon.}
\end{figure}

The vertical polarization pump beam and horizontal polarization probe beam are combined in a Glan-laser (GL) polarizer and then cross each other at an small angle of 6 mrad in the center of the cesium vapor cell. The vapor cell is 25 mm long and temperature stabilized at 112 $^{\circ}$C. The windows of the vapor cell are antireflection coated at 895 nm on both faces, resulting in a transmissivity for the far-detuned probe beam of 98\% per window. The pump beam and the probe beam are focused with waists of 560 and 300 $\mu$m (1/$e^{2}$ radius) at the crossing point, respectively, to ensure that they overlap over almost the full length of the cell. The probe beam is amplified with a power gain of about 13.7 in the FWM process, which is accompanied by the generation of a conjugate beam with a frequency of 9.2 GHz blue-shifted from the pump beam. The optical powers of the probe beam and the conjugate beam are about 204 $\mu$W and 194 $\mu$W, respectively. After the vapor cell, a Glan-Thompson (GT) polarizer with an extinction ratio of $10^{5}$:1 is used to filter out the pump beam. The amplified probe and the generated conjugate beams are directly sent into a balanced photodetector (BPD) with a quantum efficiency of 98\%. The output of the BPD is sent to a spectrum analyzer (SA), which measures the intensity-difference noise power spectrum of the twin beams. To measure the SNL, a coherent laser beam, whose power is equivalent to the total power of the probe and conjugate beams, is split into two beams using a 50:50 beam splitter and sent to the BPD.

\begin{figure}[t]
\includegraphics[width=8.5cm]{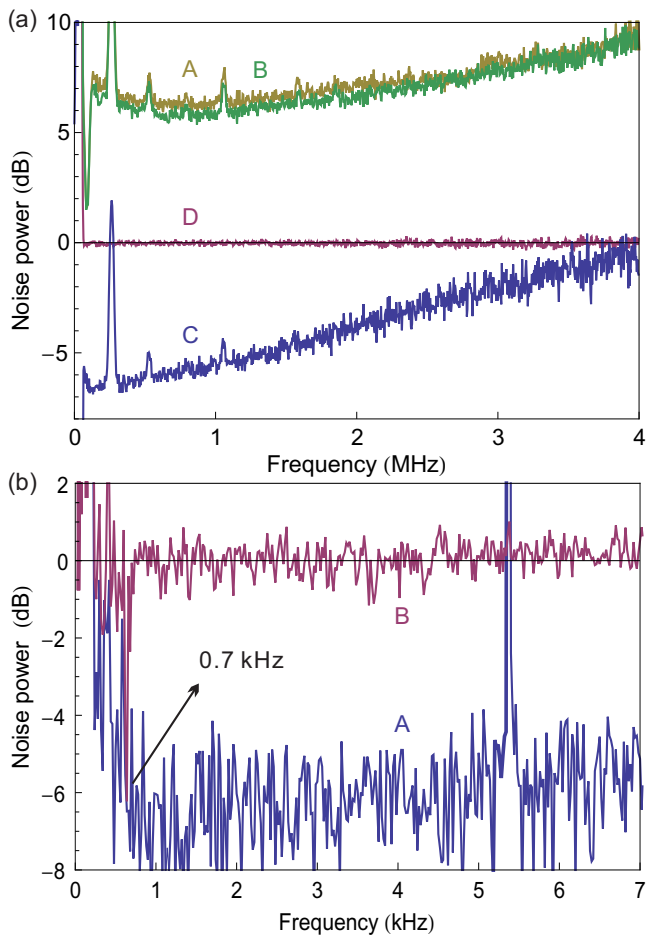}
\caption{(Color online) (a) Experimental result of intensity-difference squeezing. Noise power spectra for the probe beam ($A$), the conjugate beam ($B$) , intensity difference between the probe and the conjugate beams ($C$), and corresponding SNL ($D$). The resolution bandwidth (RBW) and video bandwidth (VBW) of the SA are 30 kHz and 300 Hz, respectively. (b) Low-frequency intensity-difference squeezing. Noise power spectra for the intensity-difference of the probe and the conjugate beams ($A$) and corresponding SNL ($B$). RBW and VBW are 10 Hz for this measurement.}
\end{figure}

\begin{figure}[tbph]
\includegraphics[width=8.5cm]{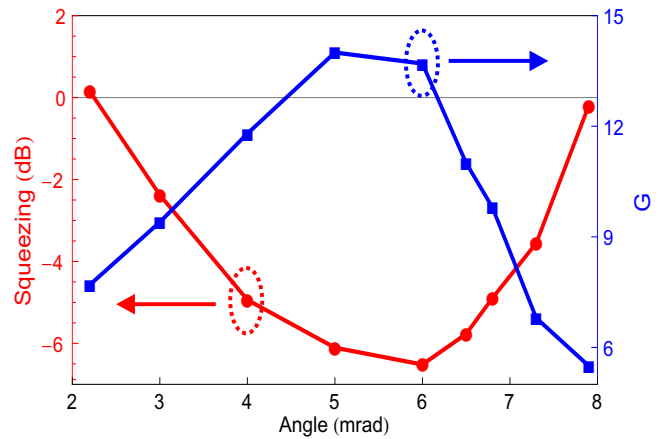}
\caption{(Color online) FWM gain (blue squares) and degree of intensity-difference squeezing (red circles) as a function of the crossing angle between the pump and probe beams.}
\end{figure}

As shown in Fig. 2(a), the red and blue curves show the transmission spectra of the scanning FP interferometer when  the EOM modulation is off and on, respectively. In both cases, the diode laser frequency is fixed at the pump frequency. With the EOM modulation on, the relative optical power in the first-order sidebands and carrier amounts to 6\%, 6\%, and 88\%, respectively. The saturated absorption spectrum (SAS) of the $^{133}$Cs $D_{1}$ line is used as the frequency reference, which is recorded by scanning the diode laser.

The temperatures of the three etalons are controlled by three individual standard proportional integral differential temperature control systems with a long-term stability of 0.005 $^{\circ}$C to make sure the peak transmission efficiency of the $-1^{st}$ order sideband of the EOM. The transmission spectrum of the three etalons is measured by scanning the diode laser [Fig. 2(b)]. We measure a free spectrum range (FSR) equal to 14.3 GHz between two transmission peaks, which is consistent with the expected value of $c/2nL$ = 14.3 GHz for the 7 mm etalon. The on-resonance transmission for the $-1^{st}$ order sideband reaches 0.85. However, for the $+1^{st}$ order sideband and the carrier frequency, we find an extinction ratio of over 40 dB.

The experimental result of quantum correlation between the probe and conjugate beams is shown in Fig. 3(a). We first record the photocurrent noise power spectra of the probe beam, conjugate beam, and intensity difference between the probe and conjugate beams, respectively (indicated as traces $A$, $B$, and $C$). All of these three traces are normalized to the corresponding SNL (trace $D$). The black straight line at 0 dB is taken as a reference and corresponds to the average value of the data points on the SNL (trace D). Trace $A$ and trace $B$ are around 6 dB above the corresponding SNL at 1 MHz, because the noise of the probe beam and conjugate beam is amplified in the FWM process. As we can see, the intensity-difference noise power of the twin beams has a maximum of 6.5 dB below the SNL with a squeezing bandwidth of 4 MHz. The large peak shown at 0.3 MHz is classical noise from our laser. This peak is eliminated perfectly on trace $D$, which shows good noise cancellation in our balanced detection system.

We next investigate the squeezing properties at audio band. Another low-noise BPD is utilized in this measurement. As shown in Fig. 3(b), the intensity-difference noise spectrum above 0.7 kHz is almost flat at 6 dB below the SNL except for a peak at 5.3 kHz. Below 0.7 kHz, the technical noise of the BPD starts to dominate, which decreases the level of squeezing. 

The observed low frequency limit for the intensity-difference squeezing is much lower than the linewidth of the laser ($\sim$ 100 kHz), which is mostly limited by frequency noise. This can be understood by two reasons as follows. Firstly, our previous work shows that the one-photon gain resonance of the FWM process in hot cesium vapor is very broad ($\sim$ 1 GHz) \cite{MaPRA}, which is much broader than that obtained in hot rubiudium vapor ($\sim$ 400 MHz) \cite{LettPRA}. As a result, the effective gain varies slowly with the one-photon detuning, i.e., the frequency of the diode laser, which avoids converting the frequency noise into amplitude noise. Secondly, the very good frequency and phase-difference between the pump and probe beams in our experiment reduces the two-photon frequency noise to minimum.

We also investigate the phase matching condition of the FWM process in hot cesium vapor. Fig. 4 shows the FWM gain and degree of intensity-difference squeezing as a function of the crossing angle between the pump and probe beams. The degree of squeezing is obtained by measuring the intensity-difference noise power of the twin beams and the corresponding SNL at 0.15 MHz and then calculating the difference of these two mean values of data points on these two curves. It clearly shows that the optimal crossing angle for the maximal degree of intensity-difference squeezing is around $\theta_{0}=6$ mard. The strong dispersion of the index of refraction for the probe beam allows the existence of 9.2 GHz frequency difference for the pump and probe beams such that the FWM process is phase matched for a nonzero crossing angle \cite{LettPRL,PhaseMatching}. The width of the squeezing dip shows the angular bandwidth of the FWM process $\Delta\theta$ is around  6 mrad, for which the pump and probe beams are quasi-phase-matched. The wide phase matching angle confirms the spatial-multi-mode characteristics of the FWM process.

In summary, we have demonstrated a compact diode-laser-pumped system for the generation of intensity-difference squeezing down to 0.7 kHz with a maximum squeezing of 6.5 dB based on the FWM process in hot cesium vapor. It is expected that the low frequency limit can be further decreased  by adding a mode cleaner on the output of the diode laser and using a BPD with lower electronic noise at audio band. Phase matching condition in the FWM process is also investigated, which confirms the multi-spatial-mode characteristics of our system. Our multi-spatial-mode sub-kHz squeezed light source near the $^{133}$Cs $D_{1}$ line may find applications involving quantum imaging, quantum metrology, and quantum information storage in electromagnetically induced transparency media. The diode-laser-pumped system would extend the possible applications for squeezing due to its low cost, ease of operation, and ease of integration.

\textbf{Funding.} This research was supported by the National Natural Science Foundation of China (Grant No. 61601270), the Key Project of the Ministry of Science and Technology of China (Grant No. 2016YFA0301402), the Applied Basic Research Program of Shanxi Province (Grant No. 201601D202006), and the Fund for Shanxi ``1331" Project Key Subjects Construction.

\end{document}